\documentclass[12pt]{iopart}
\usepackage{times}
\usepackage{graphicx}
\newcommand{\no}{\noindent}
\newcommand{\me}{\medskip\\}
\newcommand{\ma}{{\it Email:~}}
\newcommand{\mt}{{\it Tel:~}}
\newcommand{\mf}{{\it Fax:~}}
\newcommand{\mh}{{\it Home-Page:~}}

\begin{document}
\pagestyle{myheadings}

%%%%%%%%%%%%%%%%%%%%%%%%%%%%%%%%%% page 1 %%%%%%%%%%%%%%%%

\noindent
\Huge{\bf Beyond the Desert 2002}
\bigskip
\bigskip

\noindent
%\centerline{
\Large{\bf Accelerator, Non--Accelerator and Space}
\medskip

\noindent
\Large{\bf Approaches in the NEW MILLENIUM}
\bigskip
\bigskip
\bigskip

\noindent
\Large{Proceedings of the Third International Conference 
on Particle Physics Beyond the Standard Model: 
Accelerator, \protect\newline Non-Accelerator and Space Approaches,  
\protect\newline Oulu, Finland, June 2 -- 7, 2002}
\bigskip
\bigskip
\bigskip
\bigskip

\noindent
\large{Edited by H V Klapdor--Kleingrothaus}
\vspace{10.cm}

\Large{\bf Institute of Physics Publishing}
\bigskip

\Large{\bf Bristol and Philadelphia (2003)}\\

\large{ISBN 0 7503 0934 2}\\

\small{Printed in the UK by Short Run Press, Exeter}

\newpage

\normalsize

\title{Preface}
\vspace{1.cm}

	The third conference on particle physics
	beyond the Standard Model (BEYOND THE DESERT'02 - Accelerator, 
	Non-accelerator and Space Approaches) 
	was held during 2--7 June, 2002
	at the Finish town of Oulu, almost at the northern Arctic Circle.  
	It was the first of the BEYOND conference series held outside 
	Germany (CERN Courier March 2003, pp. 29-30). 
	This decision arose at the time when the University 
	of Oulu invited us to consider the Pyh\"asalmi mine as a site 
	for our GENIUS project, and to help to push 
	the development of a Finish Underground Laboratory.
	Nowadays Oulu is a modern rapidly growing university town at the 
	Gulf of Bothnia in northern Finland. 
	The conference was held at the POHTO center at Nallikari  
	in a couple of minute-walk distance to the Gulf of Bothnia 
	with a sunny beach and 3 km bike route from Oulu centre.
	It provided an intensive and pleasant atmosphere for a restricted 
	number of about 120 participants.

	Traditionally the Scientific Programme 
	of BEYOND conferences, brought into life in 1997 
	(see CERN Courier, November 1997, pp.16-18),
	 covers almost all  
	topics of modern particle physics.
	At this conference major emphasis was on
	new theoretical developments in the fields of  
	Extension of the Standard Model by means of Grand Unified and 
	SUSY theories and Extra dimensions. 
	These subjects were discussed in 
	talks of N. Mavromatos (Oxford and CERN),
	P. Nath (Boston),
	E. Ma (Riverside),
	A. Pilaftsis (Manchester), 
	B. Bajc (Lubljana), 
	H. Bech-Nielsen (DESY),	
	I. Antoniadis (CERN)
	and others. 
	M-theory and Fundamental symmetries were considered in the talks of  
	A.E. Faraggi (Minnesota), 
	M. Cvetic (Pennsylvania), 
	M. Kirchbach (Zacatecas, Mexico), 
	T. Kuo (Stony Brook), 
	Yu. Kamyshkov (Knoxville),
	T. S\"oldner (Grenoble), 
	M. Kreuz (Grenoble), 
	and M. Morita (Tokyo).
	New results of the Search for Higgses, SUSY particles, 
	R-parity violation, 
	Leptoquarks and Excited Fermions at LEP and HERA colliders
	were presented in the talks of 
	R. Nicolaidou (Saclay),  
	S. Costantini (Roma), 
	U. Katz (DESY),  
	A. Lipniacka (Stockholm) and 
	O. Yushchenko (Protvino).

	The long-standing and intriguing problem 
	of dark matter in the Universe is a permanent
	topic at any conference aiming at new physics and new phenomena. 
	From theoretical point of view
	the dark matter problem was extensively discussed by 
	D. Nanopoulos (Texas), 
	R.L. Arnowitt (Texas),  
	A. Green (Stockholm),  
 	V.A. Bednyakov (Dubna), 
	R. Viollier (Cape Town).
	Results and perspectives for Direct 
	Dark Matter Experiments with Scintillators (DAMA project)
	and Germanium detectors with big target mass 
	(GENIUS and GENIUS-TF projects 
	(see also CERN-Courier, December 1997, pp.19)) were presented by 
	R. Bernabei (Roma) and 
	I.V. Krivosheina (Heidelberg and Nizhnij Novgorod).
	Today only these two experiments
	are in principle able to see (DAMA already may have seen) 
	a positive signal 
	from dark matter particle interaction with 
	target nuclei by means of seasonal modulation.
	Other experiments (for example with sophisticated
	cryogenic detectors and ionisation-to-heat discriminations)
	due to the very small detecting mass are at present unable to notice
	such modulation signature of WIMP interactions.
	From the talk of N. Sugiyama (Tokio) ---
	"Cosmic Microwave Background: 
	A New Tool for Cosmology and Fundamental Physics"
	it was evident that 
	an unexpectedly huge amount of fundamental information can be  
	extracted from current research into cosmic microwave background. 
	This is just one example showing that 
	astrophysical data are inevitable nowadays for
	modern particle physics.
	
	Astrophysical investigations are tightly connected with	
	the exciting question of neutrino properties.
	Cosmic high energy neutrinos
	can interact with relic neutrinos 
	producing Z-Bursts which could explain the mysterious 
	origin of extremely 
	High Energy Cosmic Rays (S. Katz, Eotvos Hungary).
	Excitingly this mechanism requires the neutrino mass 
	to be in the range 
	0.02--2.2 eV, which intriguingly fits with recent results 
	obtained from neutrinoless double beta decay of Germanium
	in the Heidelberg-Moscow experiment. 
	Neutrinos from Supernova are also in the field
	of current theoretical interest and investigations
	(A. Yu. Smirnov, Trieste and Moscow). %: "Supernova Neutrinos" 

	Undoubtedly central topic of this conference was neutrino physics.
	Modern understanding of and a general view on 
	Neutrino Masses and Mixings was 
	described by R.N. Mohapatra (Maryland).
	Consequences of the SNO Neutral Current Rate for 
	resolving the Solar Neutrino Puzzle,
	were discussed by S. Choubey (UK). 
	The Standard Solar Model and 
	modern experimental hints for an elemental composition of the Sun, 
        radically different from the present usual assumptions  
        were discussed by O.K. Manuel (Missouri).

	Extended discussion of the experimental achievements
	in solar and atmospheric neutrino oscillation experiments
	included 
	first of all the Sudbury Neutrino Observatory (M. Dragowsky)
	and its results from recent analysis with 
	pure heavy water target. SNO performed first 
	measurements of the total active neutrino flux
	and claims evidence for neutrino flavour transformation
	at a 5.3 sigma level.
	Global MSW analysis favours the Large Mixing Angle (LMA) region.
 
	The status and prospects for neutrino 
        oscillation experiments KamLAND, K2K,  
	Super-Kamiokande and new facilities like neutrino factories and 
	JHF-SK project 
	were presented and discussed.
	For example, 
	KamLAND (presented by F. Suekane from Tohoku Univ, Japan) 
	is a very long baseline reactor neutrino oscillation experiment
	(with 1000 t liquid scintillator detector),
	which is able to directly test the MSW-LMA solution only with half-year
	of data and to determine the oscillation parameters with very high 
	accuracy if the LMA case is true.
	The experiment started data taking in 2002 and first
	neutrino events are successfully recorded.  	
	Rebuilding of the Super-Kamiokande detector was started in 2002
	and full reconstruction is expected before 2007
	(T. Kajita, Tokyo).
	The physics Potential and Status of the second generation
	proton decay and Neutrino experiment ICARUS 
	(Imaging Cosmic And Rare Underground Signals) in the 
	Gran Sasso Laboratory (Italy) 
	were discussed by 
	F. Mauri (Pavia) and I. Gil-Botella (Zurich).

	An exciting problem is the nature of neutrinos.
	Are these most mysterious objects Dirac or Majorana
	particles and which are their masses. 
	One of the best tool to find the answer is neutrinoless
	double beta decay.
	The evidence for observation of neutrinoless
	double beta decay (CERN-Courier Vol. 42 (number 2) 2002) 
	of the isotope Ge-76 claimed by the team 
	of H.V. Klapdor-Kleingrothaus (MPI-Kernphysik, Heidelberg)
	on the basis of the unique 
	data of the Heidelberg-Moscow collaboration has huge resonance among
	scientifists.
	This result inevitably took a central part in 
	the discussions at this conference. 
	The for this analysis crucial  
	very accurate data on the Q-Value of 
	the $^{76}{Ge}$ Double Beta-Decay 
	Determined from Accurate Mass Measurements in a Penning Trap  
	were given in the talk by I. Bergstrom (Stockholm). 
	The spokesperson of the HEIDELBERG-MOSCOW
	collaboration H.V. Klapdor-Kleingrothaus then outlined the 
	present evidence for 
	neutrinoless beta decay as well as     
	the general Future for Double Beta Experiments. 
	The collaboration 
	fixes the  effective neutrino mass
	in the region of 0.05--0.84 eV (95 \% C.L.), 
	which is meanwhile supported by independent information 
	from CMB (WMAP) and others, and by theory.
	A highly interesting new theoretical 
	conception concerning massive Majorana Particles
	was presented by D.V. Ahluwalia (Zacatecas, Mexico) who possible 
	its consequences for the structure of space-time.
	
	The important question of nuclear matrix elements 
	for double beta decay was described thoroughly 
	by F. Simkovic (Bratislava). 
	It was shown that transitions 
	to different excited daughter states could help 
	to distinguish between different mechanisms triggering 
	the neutrinoless beta decay process.
	Important new Constraints on Neutrino Mixing Parameters 
	following from the Observation of Neutrinoless Double Beta Decay
	of the HEIDELBERG-MOSCOW collaboration were also discussed by 
	H. Sugiyama (Tokyo).

 	Doubtlessly the Conference has made a remarkable  contribution  
	to fruitful exchange of ideas between the physicists  
	working in particle physics, nuclear physics and cosmology. 
	More people now believe that neutrinos at extremely 
	low energies as well as at extremely high energies 
	are the particles which can 
	supply us with new exciting discoveries in the future. 
	
	We thank all colleaques who have contributed 
	to the success of this meeting by their excellent talks 
	and lively discussions. We thank our colleague 
	Y. Peltoniemi and his team from the University of Oulu 
	for the splendid local organization of the conference. 
	We are grateful in particular also to the Local Conference 
	Secretary, Ms. Birgitta Brusila, for
	her professional and always very friendly management 
	during all the time. 

	I also would like to thank at this point, as Chairman 
	of this conference, those people, who made our visit 
	of the Pyh\"asalmi mine in February 2001 an exciting experience: 
	Jorma Kangas, at that time director of the CUPP project 
	at the University Oulu, 
	and Y. Peltoniemi, present director of the CUPP project, 
	Dr. Partti Kokkonen, Vice-Governor from the provincial 
	Goverment at Oulu,  
	but in particular Mr. Pasi Vallivaara  
	(Mayor of Pyh\"asalmi), Jukka Tikanmaki and their coworkers, 
	and the highly 
	accomodating representants of the Outokumpu mining company: 
	Timo M\"aki, chief geologist 
	and other members of the highly efficient staff.

	We are, in particular, indebted to the Scientific Secretary 
	of the conference, Dr. I.V. Krivosheina 
	(Nishnij Novgorod/Heidelberg)
	for her enthusiastic and highly efficient help 
	in scientifically organizing this conference 
	and in preparing these Proceedings.
	
\vspace{.5cm}
\hspace{7.cm}\mbox{\bf Hans Volker Klapdor-Kleingrothaus}

\hspace{7.cm}\mbox{	Conference Chairman}

\hspace{7.cm}\mbox{	Heidelberg, May 2003}

\newpage

\normalsize
\begin{tabular}{p{11.6cm}r}
\title{Contents}       &                     \\
&\\
&\\

\end{tabular}

\begin{tabular}{p{11.6cm}r}
{\bf Preface}       &                     xxx\\
&\\
{\bf 1. Grand Unified Theories and Beyond,  
SUSY/SUGRA Phenomenology}	&	1\\
\\
Quantum gravity, cosmology, (Liouville) strings and Lorentz\\ invariance       &                     \\ 
{\it N E Mavromatos}   &                     3\\
\\
CP violation in SUSY, strings and branes       &                    \\ 
{\it I Tarek and P Nath} &                     29\\  
&\\

{\bf 2. Strings}	&	39\\
\\
String and D-brane physics at low energy       &                   \\ 
{\it I Antoniadis}      &                     41             \\
&\\
Supersymmetric particle physics from intersecting D-branes       &      \\
{\it M Cvetic}       &                     57\\ 
\\ 
Phenomenological aspects of M--theory       &           \\ 
{\it A E Faraggi}   &                     67    \\
\\ 
 
{\bf 3. Extension of the Standard Model}	&	93 \\
\\
Oases in the desert: three new proposals       &           \\ 
{\it E Ma}     &                     95 \\
\\ 
Spin 1/2 bosons etc. in a theory with Lorentz violation       &       \\ 
{\it R E Allen}    &                     107  \\
\\ 
Minimal nonminimal supersymmetric standard model       &            \\ 
{\it C Panagiotakopoulos and A Pilaftsis}  &                     123     \\
\\ 
Minimal supersymmetric SU(5) theory and proton decay:\\
 where do we stand?       &                \\
{\it B Bajc, P F Perez and G Senjanovi\'c}   &                     131   \\ 
&\\
\end{tabular}

\begin{tabular}{p{11.6cm}r}

{\bf 4. Fundamental Symmetries, Properties of Space Time}	& 141\\
\\
Evidence for Majorana neutrinos: dawn of a new era in spacetime\\ 
structure       &           \\ 
{\it D V Ahluwalia}   &                     143   \\
&\\
Time-reversal violation in beta decay&            \\ 
{\it P Herczeg}     &                     161 \\
&\\
Electroweak mixing and the generation of massive Gauge Bosons       &    \\
{\it T E Ward}     &                     171 \\
\\ 
Spacetime structure of massive gravitinos       &        \\
{\it M Kirchbach and D V Ahluwalia}   &                     181   \\ 
\\ 

{\bf 5. Double Beta Decay}	&	 195\\
\\
On the $Q$-values of the tritium beta-decay and 
the double\\ beta-decay of $^{76}{Ge}$        &          \\ 
{\it I Bergstr\"om et al.}    &                     197     \\
&\\ 
 Further support of evidence for neutrinoless double beta decay 
&                    \\
{\it H V Klapdor-Klein\-grot\-haus}   &                     215\\
&\\ 
Neutrinoless double beta decay matrix elements       &           \\ 
{\it F \v Simkovic}       &                     241     \\
&\\
Consequences of neutrinoless double beta decay 
&                    \\
{\it H V Klapdor-Klein\-grot\-haus and U Sarkar}   &                     253\\
\\ 
{\bf 6. Solar and Atmosphere Neutrinos}	&	271 \\
\\
Neutrino flavor transformation observed by the Sudbury Neutrino\\
 Observatory       &               \\ 
{\it M R Dragowsky (for the SNO Collaboration)}       &         273       \\
\\ 
SNO and the solar neutrino problem       &         \\ 
{\it S Choubey et al.}       &                     291        \\
\\ 
The Standard Solar Model vs. experimental observations       &     \\ 
{\it O K Manuel}       &                 307         \\
\\ 
Status of Super-Kamiokande and the JHF-Kamioka neutrino project       &    \\ 
 {\it T Kajita (for the Super-Kamiokande and JHF-Kamioka collaborations)}       &          317
                 \\
\\
Present status of the K2K experiment       &             \\ 
{\it Y Oyama (for K2K Collaboration)}       &             331        
\\ 
\end{tabular}

\begin{tabular}{p{11.6cm}r}

Present status of the KAMLAND experiment       &            \\ 
{\it F Suekane (for the KamLAND collaboration)}  
     &                345            \\
\\  
Atmospheric neutrino oscillations with the MACRO detector       &       \\ 
{\it M Giorgini (for the MACRO collaboration)}       &         353         \\
\\
Status of the ICARUS experiment       &             \\ 
{\it F Mauri (on behalf of the ICARUS Collaboration)}       &     365      \\
\\
Physics potential of the ICARUS experiment       &           \\ 
{\it I Gil-Botella (for the ICARUS collaboration)}       &     373           \\
\\  

{\bf 7. Neutrino Mass Matrix}	&	389 \\
\\
Do Neutrino oscillations indicate a horizontal SU(2) \\
symmetry for leptons?       &                \\ 
{\it R Kuchimanchi and R N Mohapatra}       &               391          \\
\\ 
Three-neutrino MSW effect and the Lehmann mass matrix       &          \\ 
{\it P Osland}       &                 403         \\
\\ 
Constraints on neutrino parameters by neutrinoless 
double beta decay\\ experiments       &             \\ 
{\it H Sugiyama}       &                  409          \\
\\ 
{\bf 8. Neutrino Factories}	&	417\\
\\
Neutrino factories --- physics potential and present status     &\\ 
{\it O Yasuda}       &                   419       \\
\\ 
 
{\bf 9. Dark Matter, Dark Energy}	&	435 \\
\\
SUSY dark matter: closing the parameter space       &        \\
{\it R Arnowitt and B Dutta}       &                   437  \\
 &\\
Comparison of coannihilation effects in low-energy MSSM       &         \\ 
{\it V A Bednyakov}      &                   451   \\
&\\ 
Updating the constraints to CMSSM from cosmology and\\
 accelerator experiments       &               \\ 
{\it A B Lahanas, D V Nanopoulos and V S Spanos}    &           461 \\
\\ 
Interpreting data from WIMP direct detection experiments       &        \\ 
{\it A M Green}    &                  473     \\
\\ 
Searching for the dark universe by the DAMA experiment       &            \\ 
{\it R Bernabei et al.}  &                   485       \\
\\

\end{tabular}

\begin{tabular}{p{11.6cm}r}

 Search for cold dark matter and solar neutrinos 
	with GENIUS\\ and GENIUS-TF      &        \\
{\it H V Klapdor-Kleingrothaus and I V Krivosheina}   &             499 \\
\\
Sterile neutrino dark matter at the center and in the halo of the Galaxy       &               \\ 
{\it N Bili\'c et al.}     &                     515 \\
\\
{\bf 10. Cosmic Microwave Background}	&	529\\
\\
Cosmic Microwave Background: a new tool for cosmology and\\ fundamental 
physics       &         531    \\
{\it N Sugiyama}       &                      \\ 
\\ 
{\bf 11. High Energy Neutrinos, Cosmic Rays, Supernova Neutrinos}	& 541\\
\\
High energy neutrinos and tau airshowers in standard and new physics     & \\ 
{\it D Fargion}       &     543        \\
\\  
Z-Burst scenario for the highest energy cosmic rays     &    \\ 
{\it Z Fodor, S D Katz and A Ringwald}       &     567     \\
\\ 
Supernova neutrinos and neutrino mass spectrum     &     \\ 
{\it A Yu Smirnov}       &     589  \\
\\ 
\end{tabular}

\begin{tabular}{p{11.6cm}r}

{\bf 12. Search for New Physics at Colliders}	&  607\\
\\
Final achievements on Higgs boson searches at the LEP collider     &    \\ 
{\it R Nikolaidou}       &        609    \\
\\
Search for R-parity violation at LEP     &      \\ 
{\it S Costantini}       &      623      \\
\\ 
Searches for leptoquarks and excited fermions at HERA     &      \\ 
{\it U Katz (for the H1 and ZEUS Collaborations)}       &   635    \\
\\ 
Understanding   SUSY limits from LEP     &       \\ 
{\it A Lipniacka}       &        649     \\
\\
Rare processes and anomalous gauge couplings at LEP-2     &    \\ 
 {\it O Iouchtchenko}       &     665     \\
\\  

\end{tabular}

\begin{tabular}{p{11.6cm}r}

{\bf 13. Nuclear and Weak Interaction}	& 681\\
\\
Nuclear interaction in a renormalization group approach     &   \\ 
{\it T T S Kuo et al. }       &          683   \\
\\ 
Recent investigation of the $G$-parity irregular weak nucleon 
current\\ in $\beta$ decays of spin aligned $^{12}$B   and $^{12}$N     & \\ 
{\it K Minamisono et al.}       &       695 \\
\\ 
{\bf 14. Neutron Oscillations and Neutron Decay} & 701\\
\\
The Measurement of neutron decay parameters with the
    spectrometer\\ PERKEO II     &    \\
{\it M Kreuz and H Abele}       &       703      \\ 
\\ 
New limit on T Violation in free neutron decay     &  \\ 
{\it T S\"oldner et al.}       &        709           \\
&\\ 

\end{tabular}

\vspace{1.cm}

\begin{tabular}{p{11.6cm}r}

{\bf List of Participants}
     &       717\\ 
&\\ 

{\bf Author Index}     &       733\\ 
&\\ 

\end{tabular}

%////////////////////////////////
%/////////////////////////////////////////////
%///////////////////////////////

\newpage

\no
{\Large\bf List of Participants}
\medskip
\medskip

\normalsize

\no
\textbf{Ahluwalia, Dharamvir V.}\\
Department of Mathematics\\ 
Ap. Postal C-600\\
Autonomous University of Zacatecas (UAZ)\\
Zacatecas, ZAC 98060\\
MEXICO\\
\mt +52 492 92 3 9407 Ext. 211, 212\\
\mf +52 492 92 2 9975\\
\ma ahluwalia@heritage.reduaz.mx\\
\mh $http://phases.reduaz.mx/$\\ 
\me

\no
\textbf{Allen, Roland E.}\\
Department of Physics\\ 
Texas A\&M University, College Station\\ 
Texas 77843-4242\\
USA\\
\mt (979) 845-4341\\
\mf (979) 845-2590\\
\ma allen@tamu.edu\\
\mh\\ $http://faculty.physics.tamu.edu/allen$\\
\me

\no
\textbf{Antoniadis, Ignatios}\\
CERN - Theory Division\\
1211 Geneva 23\\
SWITZERLAND\\
\mt +41-22-7673201\\
\mf +41-22-7673850\\
\ma Ignatios.Antoniadis@cern.ch\\
\me

\no
\textbf{Arnowitt, Richard}\\
Department of Physics\\
Center for Theoretical Physics\\
Texas A\&M University\\
College Station, TX 77843-4242\\
USA\\
\mt (979) 845-7746\\
\mf (979) 845-2590\\
\ma arnowitt@physics.tamu.edu\\
\mh $http://www.physics.tamu.edu/\\people/person.html/Arnowitt,Richard$\\

\pagebreak

\no
\textbf{Aaltonen, Jari}\\
Gamba Oy, Hakamaantie 18\\
 90400 Kempele\\
 FINLAND\\
\ma jariaaltonen@hotmail.com\\
\me

\no
\textbf{Bajc, Borut}\\
Institut Jozef Stefan\\
Jamova 39, P.O. Box 3000\\
1001 Ljubljana\\
SLOVENIA\\
\mt 386-1-477-3781\\
\mf 386-1-2519-385\\
\ma borut.bajc@ijs.si\\
\mh $http://www.physics.nyu.edu/\\people/bajc.borut.html$\\
\me

\no
\textbf{Bednyakov, Vadim}\\
Laboratory of Nuclear Problems, JINR\\
Jolio Kuri street, 6\\
141 980 Dubna\\
RUSSIA\\
\mt 007 09621 65263\\
\mf 007 09621 666666\\
\ma bedny@nusun.jinr.ru\\
\mh $http://nuweb.jinr.dubna.su/~bedny/$\\
\me

\no
\textbf{Bech, Nielsen Holger}\\
Niels Bohr Institut\\
 Blegdamsvej 17-21\\
 DK-21100 Cobenhagen\\
 DANMARK\\
or DESY\\
 Teorieabteilung 2a, 404\\
 Notkestrasse 85\\
 D-22603 Hamburg\\
GERMANY\\
\ma hbech@alf.nbi.dk\\
\me

\pagebreak

\no
\textbf{Bergstr\"om, Ingmar}\\
Manne Siegbahn Laboratory\\
Stockholm University\\
S-10405\\
Stockholm\\
SWEDEN\\
\mt 8-161042\\
\mf 8-15 86 74\\
\ma Bergstrom@msi.se\\
\mh \\
\me

\no
\textbf{Bernabei, Rita}\\
Dipartamento di Fisica and\\
INFN-Sezione Roma 2\\
Universita di Roma "Tor Vergata"\\
Via della Ricerca Scientifica 1\\
I-00133 Roma\\
ITALY\\
\mt +39-06-72594542\\
\mf +39-06-72594542\\
\ma bernabei@roma2.infn.it\\
\mh $http://lgxserver.uniba.it/\\lei/rassegna/bernabei.htm$\\
\me

\no
\textbf{Botella, Ines Gil}\\
Institut f\"ur Teilchenphysik\\
 ETH-Honggerberg\\
        CH-8093 Zurich\\
 SWITZERLAND\\
\mt +41 1 633 38 74\\
\mf +41 1 633 12 33\\
\ma Ines.Gil.Botella@cern.ch\\
\mh $http://www.ipp.phys.ethz.ch/\\whoiswho/?document=1280$\\
\me

\no
\textbf{Brusila, Birgitta}\\
CUPP project, Box 3000\\
 FIN-90014 University of Oulu\\
Oulu\\
FINLAND\\
\mt +358-400-686 393\\
\mf +358-8-553 1353\\
\ma birgitta.brusila@pyhajarvi.fi\\
\mh $http://cupp.oulu.fi$
\me

\pagebreak

\no
\textbf{Dietz, Alexander}\\
Max-Planck-Institut f\"ur Kernphysik\\
P.O. Box 103980\\
D-69029 Heidelberg\\
GERMANY\\
\mt  +49-(0)6221-516-259\\
\mf +49-(0)6221-516-540\\
\ma  adietz@mickey.mpi-hd.mpg.de\\

\noindent
\textbf{Choubey, Sandhya}\\
Department of Physics and Astronomy\\
University of Southampton, Highfield\\
Southampton SO17 1BJ\\
UK\\
\mt +44-23-80592084\\
\mf +44-23-80593910\\
\ma sandhya@hep.phys.soton.ac.uk\\
\me

\noindent
\textbf{Costantini, Silvia}\\
c/o Dipartimento di Fisica, gruppo CMS\\
Universita` di Roma "La Sapienza"\\
Piazzale Aldo Moro 2\\
I-00185 Roma\\
ITALY\\
\mt 0039-06-4451642\\
\mf 0039-06-4453829\\
\ma Silvia.Costantini@cern.ch\\
\me

\noindent
\textbf{Cvetic, Mirjam}\\
Department of Physics and Astronomy\\
Rutgers University, Piscataway\\
NJ 08855-0849\\
USA\\
\mt 898-8153\\
\ma  cvetic@cvetic.hep.upenn.edu\\
\mh $http://\\dept.physics.upenn.edu/facultyinfo/cvetic.html$\\
\me

\noindent
\textbf{Dragowsky, Michael}\\
MS H803, PO Box 1663\\
Los Alamos National Laboratory\\
Los Alamos, NM 87545\\ 
USA\\
\mt 1-505-6674461\\
\mf 1-505-6654121\\
\ma dragowsky@lanl.gov\\
\me

\noindent
\textbf{Faraggi, Alon}\\
Theoretical Physics Department\\
University of Oxford\\
Oxford OX1 3NP\\
United Kingdom\\
\mt 01865 273960\\
\mf 01865 273947\\
\ma faraggi@thphys.ox.ac.uk\\
\mh $http://hermes.physics.ox.ac.uk/\\users/AlonFaraggi/$\\
\me

\no
\textbf{Fargion, Daniele}\\
Dipartimento di Fisica and INFN\\
Universita di Roma "La Sapienza"\\
Piazzale Aldo Moro 2\\
I-00185 Roma\\
ITALY\\
\mt +39 0649914287\\
\mf +39 064957697\\
\ma  Daniele.Fargion@roma1.infn.it\\
\mh $http://www.phys.uniroma1.it/\\DOCS/TEO/people/fargion.html$\\
\me

\noindent
\textbf{Giorgini, Miriam}\\
Universita' di Bologna and INFN\\
V.le Berti Pichat 6/2\\
I-40127 Bologna\\
ITALY\\
\mt +39-051-2095235\\
\mf +39-051-2095269\\
\ma miriam.giorgini@bo.infn.it\\
\me

\noindent
\textbf{Goradia Shantilal G.}\\
Cook Nuclear Plant\\
AEP, Bridgman\\
Michigan\\
USA\\
\ma shantilal.goradia.1@nd.edu\\
\me

\pagebreak
\noindent
\textbf{Green, Anne}\\
Physics Department\\
  Stockholm University\\
  SCFAB\\
S-106 91, Stockholm\\
SWEDEN \\
\mt +46 8 5537 8730\\
\mf +46 8 5537 8601\\
\ma amg@physto.se\\
\mh $http://www.physto.se/~amg$\\
\me

\noindent
\textbf{Iouchtchenko, Oleg}\\
CERN Meyrin\\
 CH-1211, Geneve 23\\
SWITZERLAND\\
\mt +41-22-767-42-55\\
\mf +41-22-782-30-84\\
\ma Oleg.Iouchtchenko@cern.ch\\
\me

\noindent
\textbf{Kajita, Takaaki}\\
Research Center for Cosmic Neutrinos\\
Institute for Cosmic Ray Research\\
 University of Tokyo\\
Kashiwa-no-ha 5-1-5, Kashiwa\\
 Chiba 277-8582\\
 JAPAN\\
\mt +81-4-7136-5104\\
\mf +81-4-7136-3126\\
\ma kajita@icrr.u-tokyo.ac.jp\\
\mh \\$http://www-sk.icrr.u-tokyo.ac.jp/~kajita$\\
\me

\no
\textbf{Kamyshkov, Yuri}\\
401 Nielsen Physics Building\\
 University of Tennessee\\
 Knoxville\\
Tennessee 37996-1200\\
 USA\\
\mt 865 974-6777\\
\mf 865 974-6777\\
\ma kamyshkov@utk.edu\\
\me

\pagebreak

\noindent
\textbf{Kirchbach, Mariana}\\
  Instituto de Fisica\\
  UASLP, Av. Manuel Nava 6\\
  Zona Universitaria\\
  San Luis Potosi\\
  S.L.P. 78240\\
  MEXICO\\
\mt +52 444 8 26 23 62\\
\mf +52  444 8 13 38 74\\
\ma mariana@ifisica.uaslp.mx\\
\mh $http://www.ifisica.uaslp.mx$\\
\me

\no
\textbf{Klapdor-Klein\-grot\-haus,\\
Hans Volker}\\
Max-Planck-Institut f\"ur Kernphysik\\
P.O. Box 103980\\
D-69029 Heidelberg\\
GERMANY\\
\mt +49-(0)6221-516-259\\
\mf +49-(0)6221-516-540\\
\ma klapdor@gustav.mpi-hd.mpg.de\\
\mh $http://www.mpi-hd.mpg.de/non\_acc/$
\me

\noindent
\textbf{Kreuz, Michael}\\
Institut Laue-Langevin\\
College III\\
6, rue Jules Horowitz, BP 156\\
38042 Grenoble Cedex 9\\
FRANCE\\
\mt +33 - 476207554\\
\mf +33 - 476207777\\
\ma kreuz@ill.fr\\
\me

\no
\textbf{Krivosheina, Irina}\\
Max-Planck-Institut f\"ur Kernphysik\\
P.O. Box 103980\\
D-69029 Heidelberg\\
GERMANY and\\
Radio-Physical Research Institute\\
(NIRFI), Bolshaja Pesherskaya 25\\
603 005 Nishnij-Novgorod\\
RUSSIA\\
\mt +49-(0)6221-516-259\\
\mf +49-(0)6221-516-540\\
\ma  irina@gustav.mpi-hd.mpg.de\\
\mh $http://www.mpi-hd.mpg.de/non\_acc/$
\me

\noindent
\textbf{Kuo, Thomas T. S.}\\
Department of Physics and Astronomy\\
 Stony Brook University\\
 Stony Brook, New York 11794\\
USA\\
\mt 631 632 8125\\
\mf 631 632 9718\\
\ma thomas.kuo@sunysb.edu\\
\mh $http://www.physics.sunysb.edu/\\Physics/images/img_fac/kuo.jpg$\\
\me

\no
\textbf{Laamanen, Jari}\\
Helsinki Institute of Physics\\
 P. O. Box 64\\
 00014 Helsingin Yliopisto\\
 FINLAND\\
\ma jari.laamanen@helsinki.fi\\
\me

\no
\textbf{Lehtola, Mika}\\
University of Oulu\\
 Department of Physical Sciences\\
 P. O. Box 3000\\
FIN-90014\\
 FINLAND\\
\ma mlehtola@paju.oulu.fi\\
\me

\noindent
\textbf{Lipniacka, Anna}\\
University of Stockholm\\
Fysikum, Alba Nova Center for Physics\\ 
Astronomy and Biotechnology\\
S-10691 Stockholm\\
SWEDEN \\
\mt + 46 8 55378663\\
\mf +46  8 553 78 60\\
\ma anna.lipniacka@physto.se\\
\mh $http://www.physto.se/~lipniack/$\\
\me

\no
\textbf{Maalampi, Jukka}\\
University of Jyv\"askyl\"a\\
 Department of Physics\\
 P. O. Box 35 (YFL)\\
40351 Jyv\"askyl\"a\\
 FINLAND\\
\ma jukka.maalampi@phys.jyu.fi\\
\mh $http://www.physics.helsinki.fi/~maalampi/$\\
\me

\noindent
\textbf{Ma, Ernest}\\
Department of Physics\\
University of California\\
3401 Watkins Dr.\\
Riverside, CA 92521\\
USA\\
\mt 909 787-5340\\
\mf 909 787-4529\\
\ma ma@phyun8.ucr.edu\\
\mh $http://physics.ucr.edu/\\People/Home/ma.html$\\
\me

\noindent
\textbf{Manuel, Oliver K.}\\
Department of Chemistry\\
University of Missouri\\
Rolla, MO 65401\\
USA\\
\mt 573 341-4420\\
\mf 573) 341-6033\\
\ma om@umr.edu\\
\mh $http://web.umr.edu/~om/$\\
\me

\no
\textbf{Manninen, Pekka}\\
University of Oulu\\
 Department of Physical Sciences\\
 P. O. Box 3000\\
FIN-90014\\
 FINLAND\\
\ma pekka.manninen@oulu.fi\\
\me

\noindent
\textbf{Mauri, Fulvio}\\
INFN - Sezione di Pavia\\
Via A.Bassi 6\\
I-27100 Pavia\\
ITALY\\
\mt +39 0382392419\\
\mf +39 0382423241\\
\ma Fulvio.Mauri@pv.infn.it\\
\mh $http://www.giofil.it/\\offline/440003B.HTM$\\
\me

\pagebreak

\no
\textbf{Mavromatos, Nick E.}\\
Department of Physics\\
Theoretical Physics\\
%University of London\\
King's College, London Strand\\ 
London WC2R 2LS\\
UNITED KINGDOM and\\
CERN, Theory Division\\
CH-1211, Geneva 23\\
SWITZERLAND\\
\ma  Nikolaos.Mavromatos@cern.ch\\
\me

\no
\textbf{Mohapatra, Rabindra Nath}\\
Physics Department\\
 University of Maryland\\
College Park\\
 MD 20742-4111\\
USA\\
\mt 301 405 3401\\
\mf 301 314 9525\\
\ma rmohapat@physics.umd.edu\\
\mh $http://www.physics.umd.edu/cal/\\colloquia/archives/fall01/mohapatra\_abstract.html$\\
\me

\no
\textbf{Morita, Masato}\\
Department of Physics\\
Josai International University\\
1 Gumyo,Togane\\
 Chiba 283-0002\\
JAPAN\\
\ma mmorita@jiu.ac.jp or\\
\ma g30293@m-unix.cc.u-tokyo.ac.jp\\
\me

\no
\textbf{Mutanen, Mikko}\\
University of Oulu\\
 Department of Physical Sciences\\
 P. O. Box 3000\\
FIN-90014\\
 FINLAND\\
\ma mikko.mutanen@oulu.fi\\
\me

\pagebreak

\no
\textbf{Mursula, Kalevi}\\
University of Oulu\\
 Department of Physical Sciences\\
 P. O. Box 3000\\
FIN-90014\\
 FINLAND\\
\ma kalevi.mursula@oulu.fi\\
\me

\no
\textbf{Nanopoulos, Dimitri}\\
Institute of Fundamental Physics\\
 Texas A\&M University\\
   College Station\\
 TX 77843-4242\\
USA\\
\mt 979 845-7790\\
\mf 979 845-2950\\
\ma dimitri@physics.tamu.edu\\
\mh $http://faculty.physics.tamu.edu/\\dimitri/home.html$\\
\me

\no
\textbf{Nath, Pran}\\
Northeastern University\\
Department of Physics\\
360 Huntington Ave.\\
 111 Dana\\
Boston, MA, 02115\\
USA\\
\mt 617 373-4669\\
\mf 617 373-2943\\
\ma nath@neu.edu\\
\mh $http://susy.lbl.gov/\\server-java/HEPPerson?3778$\\
\me

\no
\textbf{Nikolaidou, Rosy}\\
CEA-Saclay, DSM/DAPNIA/SPP\\
 91191 Gif sur Yvette Cedex\\
FRANCE\\
\ma Rosy.Nicolaidou@cern.ch\\
\mh $http://delphiwww.cern.ch/~nicolaid/$\\
\me

\pagebreak
\no
\textbf{Osland, Per}\\
Department of Physics\\
University of Bergen\\
Allegaten 55\\
N-5007 Bergen\\
 NORWAY\\
\mt 0047 55 58 27 68\\
\mf 0047 55 58 94 40\\
\ma Per.Osland@fi.uib.no\\
\mh \\
$http://www.fi.uib.no/~osland/osland\_per.html$\\
\me

\no
\textbf{Oyama, Yuichi}\\
High Energy Accelerator Research Organization (KEK)\\
Oho 1-1\
 Tsukuba Ibaraki 305-0801\\
 JAPAN\\
\mt +81-298-64-5434\\
\mf +81-298-64-7831\\
\ma yuichi.oyama@kek.jp\\
\mh $http://neutrino.kek.jp/~oyama$\\
\me

\no
\textbf{Pilaftsis, Apostolos}\\
Department of Physics and Astronomy\\
University of Manchester\\  
Manchester M13 9PL\\
 UNITED KINGDOM\\
\mt +44 161 275 4216\\
\mf +44 161 275 4218\\
\ma pilaftsi@theory.ph.man.ac.uk\\
\mh $http://pilaftsi.home.cern.ch/pilaftsi/$\\
\me

\no
\textbf{Katz, Sandor }\\
DESY Theory Group\\
Notkestrasse 85\\
22607, Hamburg\\
GERMANY\\
\mt +49-40-8998-3968\\
\mf +49-40-8998-2777\\
\ma katz@mail.desy.de\\
\me

\pagebreak

\no
\textbf{Katz, Uli}\\
Univ. Erlangen-N\"urnberg\\ 
Phys. Institut I\\
Erwin-Rommel-Str. 1 \\ 
91058 Erlangen \\
or  DESY\\
Notkestr. 85\\
22607 Hamburg\\
GERMANY\\  
\mt +49-9131-85/27072\\
\mf +49-9131-85/28774 \\
\ma katz@physik.uni-erlangen.de \\
\mh $http://\\zerla1.physik.uni-erlangen.de/~katz$\\
\me

\no
\textbf{Peltoniemi, Juha}\\
CUPP project\\
 Box 3000\\
 FIN-90014 University of Oulu\\
Oulu\\
FINLAND\\
\mt +358-400-686 393\\
\mf +358-8-553 1353\\
\ma juha.peltoniemi@oulu.fi\\
\mh $http://cupp.oulu.fi/juha$\\
\me

\no
\textbf{Rimsky-Korsakov, Alexander}\\
V. G. Khlopin Radium Institute\\
 28 2-nd Murinsiy Ave.\\
 St. Petersburg\\
194021\\
RUSSIA\\
\ma ark@ri.spb.su\\
\me

\no
\textbf{Saarni, Jorma}\\
Utele-Instituutti\\
 Lentokatu 2, Pilot Business Park\\
 90460 Oulunsalo\\ 
 FINLAND\\
\ma jorma.saarni@utele.net\\
\me

\pagebreak

\no
\textbf{Sarkamo, Juho}\\
University of Oulu\\
 Department of Physical Sciences\\
 P. O. Box 3000\\
FIN-90014 University of Oulu\\
 FINLAND\\
\ma jsarkamo@paju.oulu.fi\\
\me

\no
\textbf{Shen, Changquan}\\
Institute of High Energy Physics\\
 P. O. Box 913 (3)\\
 100039 Beijing\\ 
CHINA\\
\ma shencq@aspc02.ihep.ac.cn\\
\me

\no
\textbf{Simkovic, Fedor}\\
Department of Nuclear Physics\\
Faculty of Mathematics, Physics and Informatics\\
Comenius University\\
Mlynska dolina, pav. F1\\
SK-842 48 Bratislava\\
SLOVAKIA\\
\mt ++421 2 602 95 455\\
\mf ++421 2 654 12 305\\
\ma simkovic@raptor.dnp.fmph.uniba.sk\\
\me

\no
\textbf{Sinclair, David}\\
College of Natural Science\\
 Carleton University\\ 
1125 Colonel By Drive\\
Ottawa, K1S 5B6\\
CANADA\\
\ma sinclair@physics.carleton.ca\\
\me

\no
\textbf{Smirnov, Alexei}\\
The Abdus Salam ICTP\\
 Strada Costiera 11\\
 34100 Trieste\\
ITALY\\
Institute for Nuclear Research\\
 RAS, Moscow\\
 RUSSIA\\
\ma smirnov@ictp.trieste.it\\
\me

\pagebreak

\no
\textbf{S\"oldner, Torsten}\\
Institut Laue Langevin\\
  BP 156\\
  F-38042 Grenoble Cedex 9\\
  FRANCE\\
\mt +33 (0)4 76 20 70 92\\
\mf +33 (0)4 76 20 777 77\\
\ma soldner@ill.fr\\
\mh $http://\\www1.physik.tu-muenchen.de/~tsoldner/$\\
\me

\no
\textbf{Strumia, Alessandro}\\
Theoretical Physics Division\\
 CERN, CH-1211 Geneva 23\\
 SWITZERLAND\\
\ma Alessandro.Strumia@cern.ch\\
\me

\no
\textbf{Suekane, Fumihiko}\\
Research Center for Neutrino Science\\
Tohoku University\\
Sendai 980-8578\\
 JAPAN\\
\mt 81 22-217-6727\\
\mf 81 22-217-6728\\
\ma suekane@awa.tohoku.ac.jp\\
\me

\no
\textbf{Suhonen, Esko}\\
University of Oulu\\
 Department of Physical Sciences\\
 P. O. Box 3000\\
FIN-90014 University of Oulu\\
 FINLAND\\
\ma esko.suhonen@oulu.fi\\
\me

\noindent
\textbf{Sugiyama, Hiroaki}\\
Department of physics\\
 Tokyo Metropolitan University\\
1-1 Minami-Osawa, Hachioji\\
 Tokyo 192-0397\\
 JAPAN\\
\mt 81-426-77-1111 ex.3375\\
\mf 81-426-77-2483\\
\ma hiroaki@phys.metro-u.ac.jp\\
\me

\no
\textbf{Sugiyama, Naoshi}\\
Division of Theoretical Astrophysics\\
National Astronomical Observatory Japan\\
2-21-1 Osawa, Mitaka\\
 Tokyo 181-8588\\
 JAPAN\\
\mf +81-422-34-3746\\
\ma naoshi@th.nao.ac.jp\\
\mh $http://th.nao.ac.jp/~naoshi/$\\
\me

\no
\textbf{Ruppell, Timo}\\
Helsinki Institute of Physics\\
 P. O. Box 64\\
 00014 University of Helsinki\\
 FINLAND\\
\ma Timo.Ruppell@helsinki.fi\\
\me

\no
\textbf{Viollier, Raoul D.}\\
University of Cape Town\\
Institute of Theoretical Physics\\
 and Astrophysics \\     
Private Bag, Rondebosch, 7701\\
Cape Town\\
SOUTH AFRICA\\
\mt +27-21-650-3334\\
\mf 00-27-21-685-5510\\
\ma VIOLLIER@physci.uct.ac.za\\
\mh $http://www.uct.ac.za/\\depts/physics/theorygp.html$\\
\me

\no
\textbf{Ward, Thomas}\\
U.S. Department of Energy\\
MS: NA-125, Rm1J063\\
1000 Independence Ave., S.W.\\
Washington, D.C. 20585\\
 USA\\
\mt 202 586-7255\\
\mf 202 586-1966\\
\ma Thomas.Ward@nnsa.doe.gov\\
\me

\pagebreak

\no
\textbf{Yasuda, Osamu}\\
Department of Physics\\
Tokyo Metropolitan University\\
1-1 Minami-Osawa, Hachioji\\
Tokyo 192-0397\\
 JAPAN\\
\mt 81-426-77-1111 ext 3376\\
\mf 81-426-77-2483\\
\ma yasuda@phys.metro-u.ac.jp\\
\me

\end{document}